\documentclass[jgrga]{agutex2015}

\usepackage{graphicx}
\authorrunninghead{MILITZER ET AL.}

\titlerunninghead{UNDERSTANDING JUPITER'S INTERIOR }

\begin{document}

%
%

\title{Understanding Jupiter's Interior}
%
%

%
%



\author{Burkhard Militzer$^{1,2}$, Fran\c{c}ois Soubiran$^{1}$, Sean M. Wahl$^{1}$, William Hubbard$^{3}$\\[3mm] 
\small$^{1}$Department of Earth and Planetary Science, University of California, Berkeley, CA 94720, USA.\\
\small$^{2}$Department of Astronomy, University of California, Berkeley, CA 94720, USA.\\
\small$^{3}$Lunar and Planetary Laboratory, University of Arizona, Tucson, AZ 85721, USA.}




%
%




%
%


\begin{abstract}
  This article provides an overview of how models of giant planet
  interiors are constructed. We review measurements from past space
  missions that provide constraints for the interior structure of
  Jupiter. We discuss typical three-layer interior models that consist
  of a dense central core and an inner metallic and an outer molecular
  hydrogen-helium layer. These models rely heavily on experiments,
  analytical theory, and first-principle computer simulations of
  hydrogen and helium to understand their behavior up to the extreme
  pressures $\sim$10 Mbar and temperatures $\sim$10$\,$000 K. We
  review the various equations of state used in Jupiter models and
  compare them with shock wave experiments. We discuss the possibility
  of helium rain, core erosion and double diffusive convection may
  have important consequences for the structure and evolution of giant
  planets. In July 2016 the \textit{Juno} spacecraft entered orbit
  around Jupiter, promising high-precision measurements of the
  gravitational field that will allow us to test our understanding of gas
  giant interiors better than ever before.
\end{abstract}

%
%

%

\begin{article}

%
%

\section{Introduction}

In this article, we will provide a brief overview of how models for the
interiors of giant planets are put together. While much of this
discussion applies to all giant planets, this article will be focused
on Jupiter in particular. We will review results from space
missions that visited the planet earlier and then we will discuss
what we expect from the \textit{Juno} mission presently in orbit
around Jupiter.

All giant planet interior models rely on an equation of state that
describes how materials behave under the extreme pressure ($\sim$10
Mbar) and temperature ($\sim$10$\,$000$\,$K) conditions in planetary
interiors.  So in section~\ref{sec:EOS}, we compare the results from
laboratory experiments, semi-analytical EOS models and {\it ab initio}
simulations of dense hydrogen and of helium. In section~\ref{sec:HHe}, we review
experimental and theoretical predictions for the properties of
hydrogen-helium mixtures. We discuss {\it ab initio} simulations that
focused on the question whether hydrogen-helium mixtures phase separate
at high pressure, where hydrogen becomes a metallic fluid
while helium remains in an insulating state. This process may lead to
helium rain in the interior of giant planets, which has been invoked
by~\citet{Stevenson77a,Stevenson77b} to explain
Saturn's unusually large infrared emissions.

In section~\ref{sec:interior}, we compare the prediction for Jupiter's
temperature-pressure profiles and discuss various interior models. In
section~\ref{sec:conv}, we compare adiabatic and super-adiabatic
models, revisit the question of whether present day Jupiter has dense
central core and if a primordial core could be partially or fully eroded.

\section{Interior Constraints from Past Space Missions}
\label{sec:space_missions}

Over the past 43 years Jupiter has been visited by nine spacecraft. Out of
these missions the primary contributions to our understanding of Jupiter's
interior were made by the \textit{Pioneer} 10 \& 11 fly-bys, the
\textit{Voyager} 1 \& 2 fly-bys and the \textit{Galileo} orbiter.  In July
2016, the \textit{Juno} spacecraft entered a low-periapse orbit, in order to provide
the most precise measurement of Jupiter's gravitational field to date, as well
as better constraints on the composition of the outer envelope.

\begin{figure}[ht]
  \begin{center}
    \noindent\includegraphics[width=20pc]{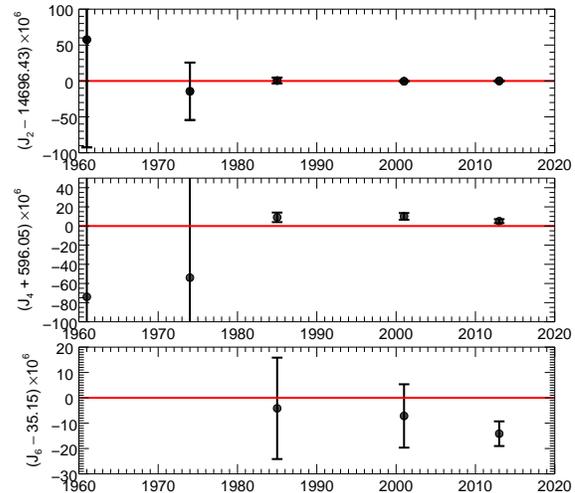}
  \end{center}
\caption{Improvement in measurements of Jupiter's first even zonal harmonics, as a function of year
(abscissa).  All $J_n$ values are normalized to $a$ = 71492 km, and referenced to theoretical
values from a recent Jupiter model \citep{HM16}, horizontal red line. }
\label{fig:jplot}
\end{figure}

The most direct constraints on Jupiter's interior structure comes from
measurements of the non-radial terms in its external gravitational
potential.  Jupiter's large surface oblateness (second only to
Saturn's in the solar system) is a visible manifestation of its rapid
rotation rate and low mean density.  Correspondingly, the planet's
external gravitational potential has large zonal harmonic coefficients
$J_n$, which can be measured by modeling their effects on the orbit of
a nearby spacecraft or natural satellite.  These coefficients are
weighted integrals over the interior density distribution $\rho({\bf
  r})$,
\begin{equation}\label{eq:J2ndef}
J_n = - \frac{2 \pi}{M a^n}  \int dr \, d\mu \; \rho({\bf r}) \; r^{n+2} \; P_n(\mu) ,
\end{equation}
where $M$ is Jupiter's mass, $a$ is a normalizing radius (usually taken to be the
equatorial radius at a pressure of 1 bar, 71492 km), $\mu = \sin L$ ($L$ is the
planetocentric latitude), $P_n(\mu)$ are Legendre polynomials, and $r$ is the radial
distance from the planet's center. To relate given values of $J_n$ to interior
structure, we assume that the planet is everywhere in hydrostatic equilibrium in its
rotating frame, and that a unique barotrope $P=P(\rho)$ relates the pressure $P$ and
the mass density $\rho$.  Thus, a model of Jupiter using a barotrope that reproduces
the external gravity terms is an acceptable one.

The $J_2$ term in the harmonic expansion is mostly Jupiter's interior's linear
response to rotation, but higher-order terms $J_n$ arise entirely from nonlinear
response, and require careful numerical modeling to properly test an assumed interior
barotrope \citep{ZT1978}.  The higher-order terms are difficult to measure at a significant
distance from Jupiter since the gravitational potential contribution from a given
zonal harmonic $J_n$ varies as $(a/r)^{n+1}$.  Prior to the first spacecraft
measurements at Jupiter in 1973, our only information about $J_n$ came from
ground-based observations of satellite motions.  Fig.~\ref{fig:jplot}  exhibits the
dramatic improvement in measurements of the $J_n$ over the last $\sim$50 years.

\begin{figure}[ht]
  \begin{center}
    \noindent\includegraphics[width=16pc]{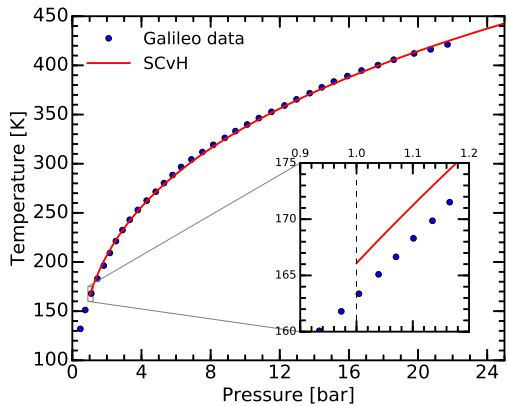}
  \end{center}
  \caption{Comparison of Galileo entry probe $T$-$P$ measurements
    \citep{GalileoData} with \citet{SC95} EOS model. For the latter
    calculation, we assumed a dry adiabat with helium mass fraction of
    0.247 a starting from a temperature of 166.1~K at 1~bar. }
\label{fig:ad}
\end{figure}

\subsection{Pioneer and Voyager Missions}

\textit{Pioneers} 10 and 11 were the first spacecraft to reach the vicinity of
Jupiter, in December 1973 and December 1974. These spacecraft executed hyperbolic
flyby orbits with periapses at $2.8 a$ and $1.6 a$ respectively, permitting an
improved determination of $J_2$ and $J_4$, but the signal from higher-order terms was
not measurable.  \textit{Voyagers} 1 and 2 executed Jupiter flybys in March 1979 and
July 1979, with periapses of $4.9 a$ and $10.1 a$ respectively.  Thus, the
\textit{Pioneer} flybys were more sensitive to terms above $J_2$
\citep{CS85}.

\subsection{Galileo Mission and Entry Probe}

The \textit{Galileo} spacecraft arrived at Jupiter in December 1995, and remained in orbit
around Jupiter for 8 years. The main \textit{Galileo} spacecraft did not make a large
contribution to improvements in measurements of Jupiter's $J_n$, although it made
important measurements of the corresponding terms in the external gravity of the
Galilean satellites.

\textit{Galileo}'s most significant contribution to the understanding
of Jupiter's interior was deploying an entry probe that measured the
structure and composition of Jupiter's near-equatorial atmosphere to a
maximum pressure of 22 bar~\citep{seiff-1998}.  The entry probe data
are fundamental as an initial condition for constraining the Jovian
interior barotrope. Conservatively estimated, the probe's 
sensors were able to measure temperature with a precision 
varying from 0.1~K at 100~K to 1~K at 500~K~\citep{Magalhaes2002}.
\citet{seiff-1998} inferred an uncertainty of less than 2\% for
probe's pressure measurements by performing an \textit{a posteriori}
calibration that was needed because the temperature exceeded the
original calibration range.  Based on these results,
\citet{seiff-1998} fitted a dry adiabatic profile with a temperature
of 166.1~K at 1~bar, which we reproduced in Fig.~\ref{fig:ad} with a
H-He isentrope derived from the \citet{SC95} EOS model. We find a good
agreement overall between the theoretical H-He isentrope and the
measured data but is a not a perfect match. For example there is
deviation of approximately 2~K at starting point of 1 bar.

The \textit{Galileo} entry probe further constrained the composition
along the barotrope, showing the outer layers of Jupiter to have a
composition that is not too different from that of the sun. The probe
measured a helium mass fraction of $Y=0.23$ and a mass fraction of
heavier elements of $Z \approx 0.017$. The heavy element component was
primarily comprised of the hydrides H$_2$O, CH$_4$, and
NH$_3$~\citep{Wong2004}. It is important to note that the probe value of $Y$ is below
the protosolar value of 0.274~\citep{Lodders03}, providing evidence of
helium sequestration in Jupiter's interior. The probe's measurement of a
strong neon depletion with respect to protosolar abundance was further
evidence of helium rain~\citep{WilsonMilitzer2010}.

\subsection{Combined Analysis from Past Missions}

The next spacecraft to visit Jupiter was Ulysses, which executed a flyby in February
1992 at $6.3 a$. Optimized to study the solar wind, Ulysses did not contribute
significantly to Jovian interior constraints. Subsequent encounters by
\textit{Cassini-Huygens} in December 2000 and by \textit{New Horizons} between
January and may 2007, did not afford opportunities to significantly improve
measurements of Jupiter's gravitational field.

During the long interval before the expected arrival of the Juno orbiter in 2016, R.
A. Jacobson \citep{Jacobson2001,Jacobson2003,Jacobson2013} has synthesized disparate
data sets including Earth-based astrometry, satellite mutual eclipses and
occultations, and satellite eclipses by Jupiter, as well as spacecraft data from
Doppler tracking, radiometric range, very-long baseline interferometry, radio
occultations, and optical navigation imaging from \textit{Pioneer} 10 \& 11,
\textit{Voyager} 1 \& 2, \textit{Ulysses}, \textit{Galileo}, and \textit{Cassini}.
More recent data points shown in Fig.~\ref{fig:jplot} reflect this work.

\subsection{Juno Mission}

The Juno spacecraft, launched in 2011 and inserted in
  Jupiter's orbit on July 4, 2016, is optimized for measurements to
constrain Jupiter's interior structure.  More than 20 polar orbits
with 14-day periods and perijoves at $\sim1.07 a$ will be devoted to
$X$- and $Ka$-band measurements of spacecraft motions in Jupiter's
gravity potential, with an expected line-of-sight velocity precision
$\sim2 \mu$m/s.  Terms in Jupiter's gravitational potential to
$\sim$$J_{10}$ should be measurable, along with Jupiter's
second-degree tidal response to its nearest large satellites.
Predicted values of Jupiter's $J_2$, $J_4$, and $J_6$ are shown in
Fig.  \ref{fig:jplot}; predictions of $J_8$ and $J_{10}$ are also
published \citep{HM16}.  Predicted values of Jupiter's tidal Love
numbers~\citep{WHM16a,WHM16b} are available for comparison with the
tidal measurements.  The precision of predictions and expected
measurements is such that relative discrepancies at the level of
$\sim10^{-7}$ would be detectable.  Gravity anomalies attributable to
Jovian interior dynamics, apart from purely hydrostatic response, may
produce a detectable signal~\citep{KG16}.

Over the expected mission lifetime of $\sim$20 months, a sufficient
arc of the angular precession of Jupiter's spin axis should be
measurable to yield a meaningful result for Jupiter's spin angular
momentum, $L=C \omega$, where $C$ is Jupiter's axial moment of inertia
and $\omega$ is the spin rate.  If the relevant $\omega$ is the
well-known and stable value for Jupiter's magnetic field, virtually
any interior model fitted to $J_2$ predicts $C = 0.26 M
a^2$~\citep{HM16}. Measurement of an $L$ significantly different from
the value implied by Jupiter's magnetic field rotation rate would
suggest differential rotation involving a substantial fraction of the
planetary mass.

The microwave radiometer (MWR) experiment on Juno will probe
abundances of the condensable gases H$_2$O, NH$_3$, and H$_2$SO$_4$ by
sounding Jupiter's deep atmosphere at six wavelengths from 1.37 to 50
cm, with sensitivity to levels at pressures ranging from $\sim$ 1 bar
to $\sim$ 100 bar~\citep{Janssen2014}. Although results from MWR may
confirm the Galileo probe value for a metallicity $Z \approx 0.017$ in
the outermost region of the jovian barotrope, a significant change in
the metallicity would change the inferred mass density of Jupiter's
outer layers with repercussions on jovian structure at deeper layers.

\begin{figure}[!ht]
  \begin{center}
\noindent\includegraphics[width=16pc]{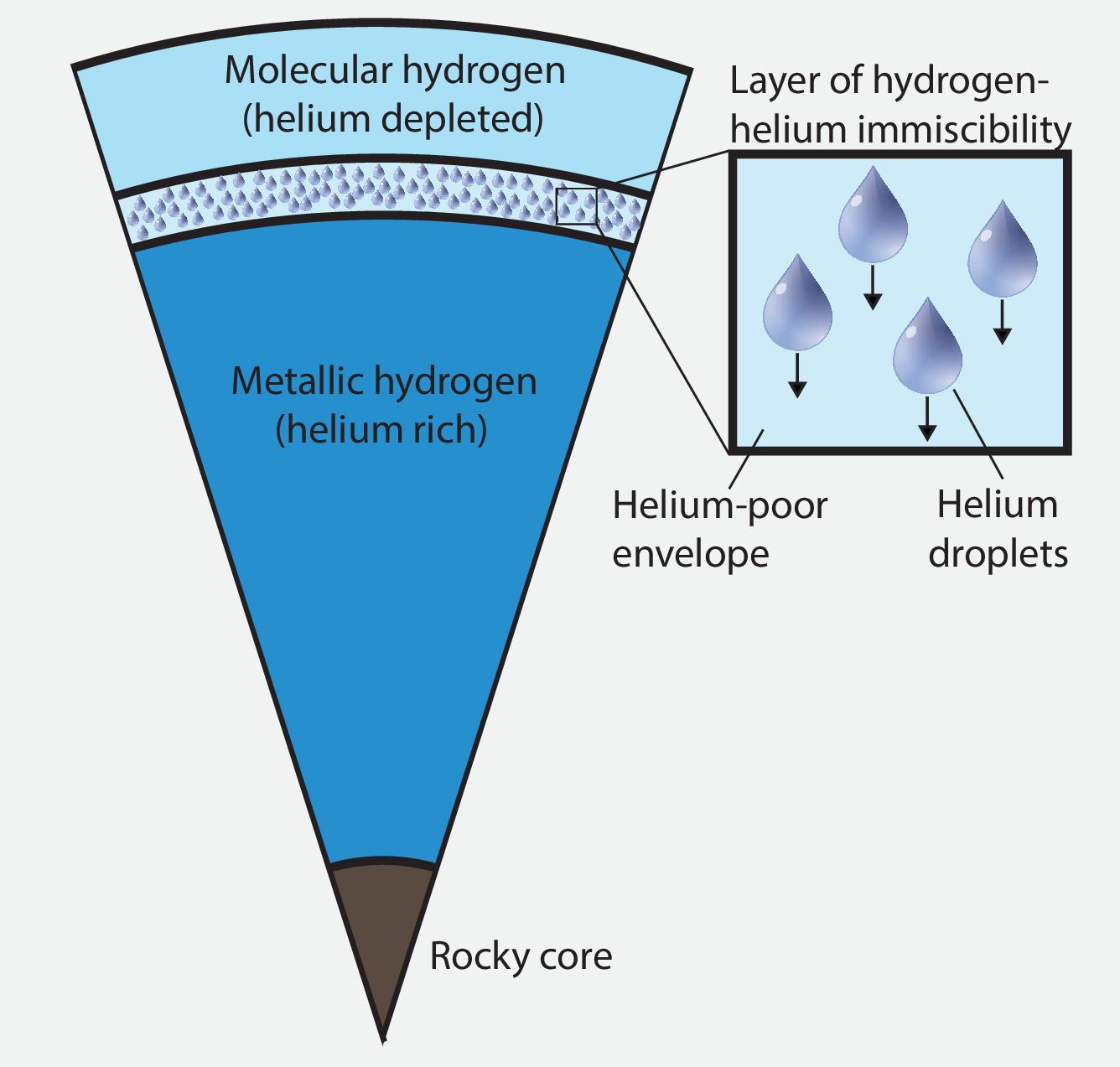}
\noindent\includegraphics[width=16pc]{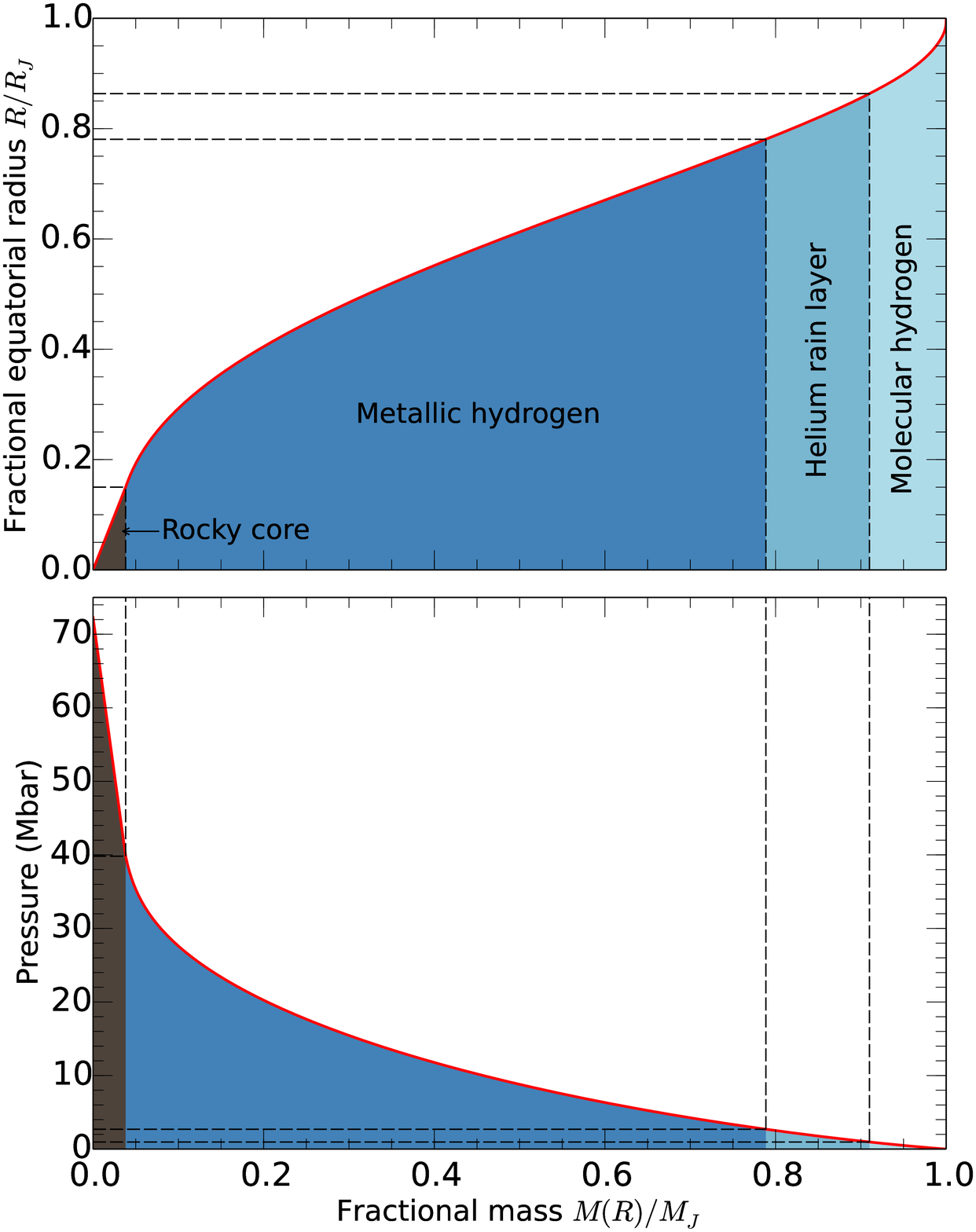}
  \end{center}
  \caption{The upper diagram shows a model of Jupiter's interior with the
    hydrogen-helium immiscibility layer. Lower two diagrams show the
    fractional radius and pressure as a function of fractional mass
    according to \citet{HM16}.}
\label{fig:our_Jupiter_model}
\end{figure}

\section{Equations of State of Hydrogen and Helium}
\label{sec:EOS}

\subsection{Theory, Simulations, and Shock Wave Experiments of Dense Hydrogen}

Hydrogen is the most common element in the universe. Hydrogen and helium
are the dominant elements in the interiors of main-sequence stars and
gas giant planets. Because of this astrophysical context, the equation
of state (EOS) of hydrogen has been studied with various methods for
many decades. Here we will review a selected set of articles that
contributed to our understanding of dense, molecular hydrogen in the
outer envelope of giant planets as well as of metallic hydrogen in
their deep interior (Fig.~\ref{fig:our_Jupiter_model}). Because Jupiter
has a strong, dipolar magnetic field, we know conducting, metallic
hydrogen must be present in its interior.

Long before laboratory experiments or {\it ab initio} computer
simulations became available, the EOS of dense hydrogen plasma was characterized with analytical free energy
models~\citep{braunes_buch} that invoke the {\em chemical picture}. In
this approach, one describes plasmas as a collection of charged ions
and electrons as well as neutral particle such as molecules and atoms.
Approximate free energy functions are derived for each species and the
chemical composition is obtained by minimizing the combined free
energy for a given pressure and temperature.

\begin{figure}[!ht]
  \begin{center}
    \noindent\includegraphics[width=16pc]{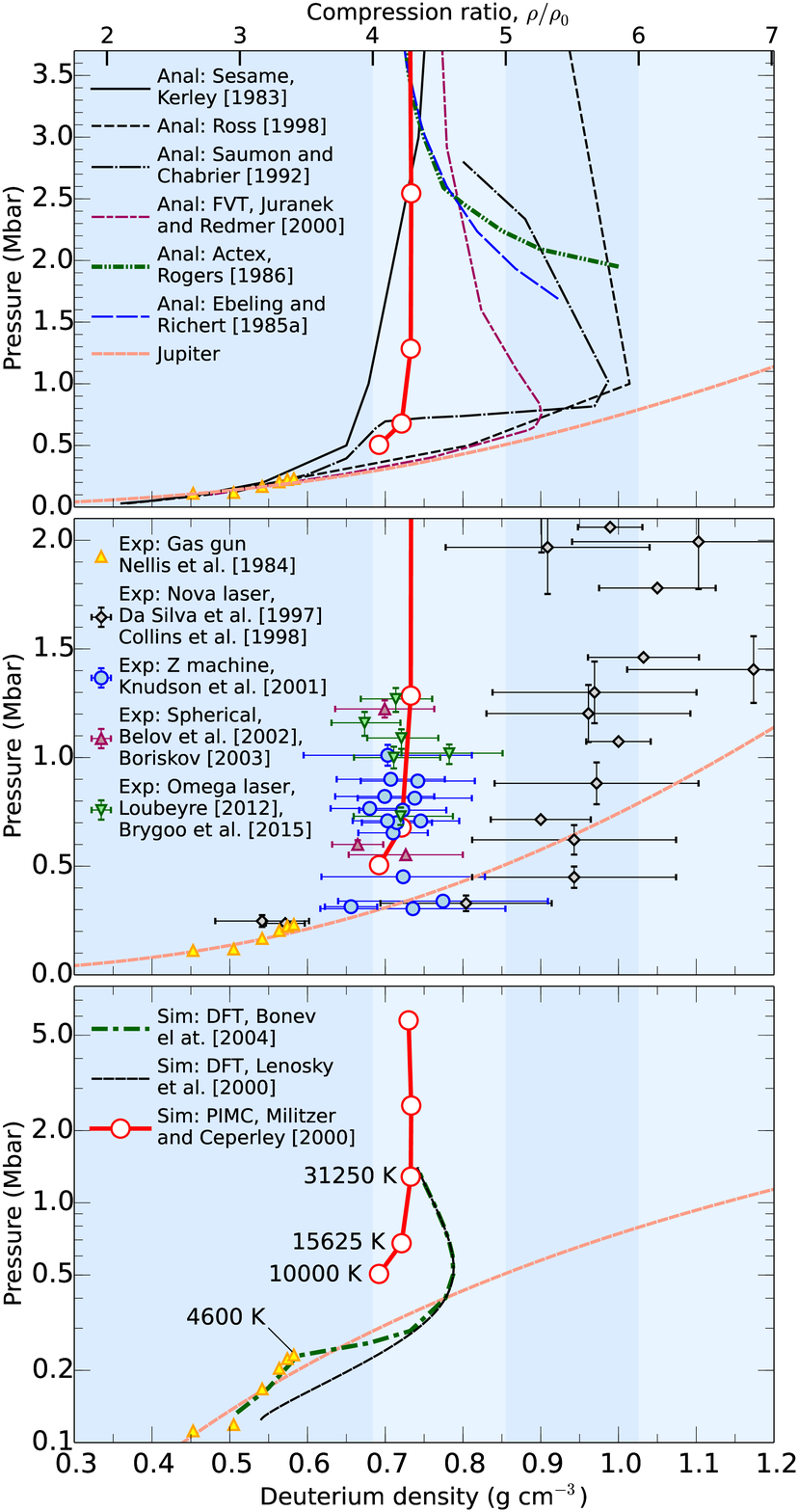}
  \end{center}
  \caption{Comparison of the deuterium shock Hugoniot curves derived
    with semi-analytical methods (upper panel: \citet{Ro86},
    \citet{ER85b}, \citet{SC92}, \citet{Ke83}, \citet{Ju00},
    \citet{Ro98}), shock wave experiments (middle panel:
    \citet{Nellis1984}, \citet{Si97}, \citet{Co98}, \citet{Kn01},
    \citet{Belov2002}, \citet{Boriskov2003}), and first-principles
    computer simulations (lower panel: \citep{MC00}, \citep{Le00},
    \citet{Bonev2004}). Changes in backgound color mark densities
    equal to 4, 5, and 6 times the initial deuterium density of
    $\rho_0=0.17$ g$\,$cm$^{-3}$.}
  \label{fig:hydrogen}
\end{figure}

Chemical models are known to work very well in regimes of weak
interaction. At low density, the ionization equilibrium can be derived
from the ideal Saha equation~\citep{Fo65}, which neglects all
interactions. Various elaborate analytical schemes have been derived
to introduce interaction effects into free energy models. Not all of
these were constructed to describe the whole high-temperature phase
diagram as done by \citet{SC92}. \citet{ER85a} studied the plasma and
the atomic regime, while models by \citet{Be99} and \citet{BN97} were
designed to describe the dissociation of molecules. The
\citet{Ro98} model was primarily developed to study the
molecular-metallic transition. One difficulty common to free energy
models is how to treat the interaction of charged and neutral
particles. Often, this is done by introducing hard-sphere radii and
additional corrections. These kinds of approximations lead to
discrepancies between various chemical models. The differences are
especially pronounced in the regime of the molecular-to-metallic
transition because of the high density and the presence of neutral and
charged species. If the derivatives of the free energy are continuous
in this regime, a gradual molecular-to-metallic transition is
predicted. If, on the other hand, the different
components of the free energy lead to discontinuous first derivatives,
a first-order transition, or plasma phase transition (PPT) is
inevitably predicted. 

The question whether such PPT exists remains controversial. Many
models have predicted a PPT with a critical point and coexistence
region of two fluids characterized by different degrees of ionization
and densities. A PPT was first placed on the hydrogen phase diagram by
\citet{La43}. First calculations have been made by \citet{No68} and
\citet{Eb73}. A number of different free energy models such as those
by \citep{SC92,Ki97,Be99} predict a PPT. The exact location of the
critical point and the coexistence region differ considerably and other
models show continuous transitions \citep{Ro98}. Since this was an
open question, \citep{SC92} provided an alternate model where they
smoothly interpolate between both regimes.

\begin{figure}[tbh]
  \begin{center}
    \noindent\includegraphics[width=16pc]{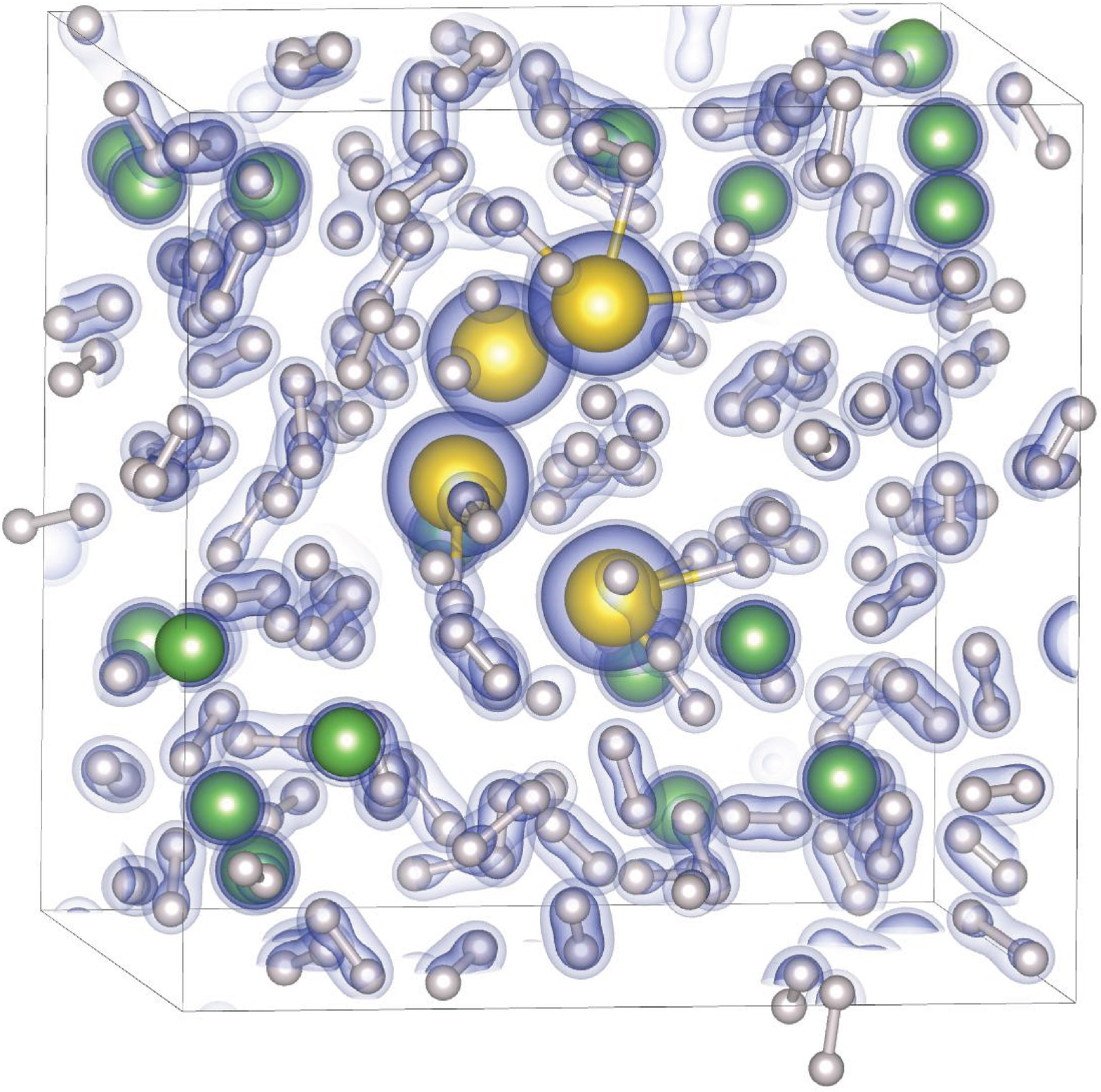}
  \end{center}
  \caption{Snapshot from a DFT-MD simulation of 220 hydrogen,
      18 helium, and 4 iron atoms that were introduced as an example of
      heavier elements~\citep{Soubiran2016}. The grey isosurfaces
      represent the density of valence electrons. Periodic boundary
    conditions were used to mimic a macroscopic system.}
  \label{fig_snapshot}
\end{figure}

Path integral Monte Carlo simulations by \citet{Ma96} showed evidence
of a first order transition in dense hydrogen. However, it was
predicted to occur at relatively low temperatures~\citep{MG06}, for
which PIMC results showed some dependence on the choice of fermion
nodes. In this temperature regime, density functional molecular
dynamics (DFT-MD) simulations work very efficiently. A snapshot from
such simulations is shown in Fig.~\ref{fig_snapshot}. Simulations
by~\citet{Vo07} and \citet{MHVTB} predicted a gradual
molecule-to-metallic transition for temperature conditions in the
interiors of Jupiter and Saturn. However, at lower temperatures ($<$
2000$\,$K), well below the giant planet interior adiabats, DFT-MD
simulations also predict a first-order
transition~\citep{Morales2010,Lorenzen2011}.

Significant progress has been made with high pressure laboratory
experiments since the reverberating shock wave measurements by
~\citet{We96} first produced hot, metallic hydrogen in the laboratory.
Still it has remained a challenge to determine whether the
molecular-to-metallic transition is of first-order. Measurements by
\citet{Fortov2007} and most recently by \citet{Knudson15} both showed
evidence of a first-order transition.  Still more work is needed to
reconcile the results from different experiments with each
other and develop a consistent theoretical framework.

The existence of a first-order molecular-to-metallic transition at sufficiently high
temperature could introduce a convective barrier into Jupiter's
interior. It would thus delay cooling of its interior and stabilize
a compositional difference between the molecular and metallic layers.
Such a difference is indeed invoked in most models for Jupiter's
interior in order to match the gravitational moment $J_4$ but the
origin of this difference remain poorly understood. We favor the
hypothesis that the convective barrier was instead introduced by helium rain, which we
will discuss in section~\ref{sec:phase_sep}. For the interior models
discussed later in this article, we will rely on DFT-MD simulations
that predict a gradual molecular-to-metallic transition for pure
hydrogen for the temperatures in Jupiter's interior.

While unanswered questions remain regarding the phase diagram of dense
hydrogen, significant progress has been made in characterizing the
shock Hugoniot curve of hydrogen, which is summarized in
Fig.~\ref{fig:hydrogen}. Shock wave experiments are the preferred
laboratory technique \citep{Ze66} to determine the equation of state
at high pressure and temperature. During such experiments, a
driving force is utilized to propel a pusher at constant velocity
$U_p$ into a material at predetermined initial conditions
($\rho_0=m/V_0, P_0, T_0$). The impact generates a planar shock wave,
which travels at the constant velocity $U_s$, where $U_s>U_p$. Behind
the shock wave, the material reaches a final state thermodynamic
equilibrium ($\rho=m/V,P,T$).  The conservation laws of mass,
momentum, and energy, $E$, across the shock interface leads to the
Rankine-Hugoniot equations~\citep{Rankine1870, Hugoniot1887,
    Hugoniot1889}, which have been discussed in detail by \citet{Ze66},
\begin{eqnarray}
\label{shock_p}
P &=& \varrho_0 U_s U_p + P_0\\
\label{shock_rho}
\rho &=& \rho_0 \frac{U_s}{U_s-U_p}\\
R(\rho,T) &=& E-E_0+\frac{1}{2}(V-V_0)(P+P_0)=0 \quad.
\label{eq:hug}
\end{eqnarray}
It is remarkable that the measurements of just $U_s$ and $U_p$ allow
for an absolute EOS measurement. The shock temperature, however, must
be determined independently. The shock Hugoniot curve in
Fig.~\ref{fig:hydrogen} emerges as collection of final states for
different $U_p$. The computation of this curve is straightforward. For
a theoretical EOS provided in term of $E(\rho,T)$ and $P(\rho,T)$, one uses
Eq.~\ref{eq:hug} to solve for $R(\rho,T)=0$ by varying $\rho$ at fixed
$T$.

In Fig.~\ref{fig:hydrogen}, we compare the experimental Hugoniot
curves with predictions from analytical techniques and first-principles
computer simulations based on PIMC and DFT-MD. Because of shock
heating, the temperature along shock Hugoniot curve rise significantly
more rapidly than that on an adiabat, which makes it difficult to
relate shock wave measurements to planetary interiors~\citep{MH08}. In
$P$-$\rho$ space, the Hugoniot curves plot at higher pressures than
Jupiter's adiabat. Nevertheless, these measurements provide invaluable
constraints for the theoretical EOS calculations.

In the upper panel of Fig.~\ref{fig:hydrogen}, we compare various
analytical EOS models. In the limits of high pressure, all curves are
expected to converge to 4-fold compression, the limiting case for a
non-relativistic gas ($\rho/\rho_0=4$). The activity expansion (ACTEX)
by ~\citet{Ro86}, the Pad\'e approximations in the chemical picture by
\citet{ER85b} and \citet{SC92} EOS model all predict compression ratio
of $\sim$5.5 at 2 Mbar and then converged to limit of 4-fold
compression at higher pressure. The Sesame model by \citet{Ke83},
which has been used frequently to simulate a variety of shock
processes, predicts a lower shock compression in comparison. The fluid
variational theory (FVT) by \citet{Ju00} yield compression ratios up
to 5.2. The linear mixing model by \citet{Ro98} stands out among all
EOS models because it predicts shock compression ratios between
5.5 and 6.0 in the entire pressure interval from 1 to 4 Mbar. 

When laser-driven shock wave experiments by \citet{Si97} and
\citet{Co98} generated megabar pressures in deuterium for the first
time, the results were surprising because they implied shock
compression ratios of $\sim6$ and therefore favored the linear mixing
model by \citet{Ro98}. These measurements sparked an intense debate in
the high-pressure community and motivated additional experimental and
theoretical work. Points of concern were that neither the PIMC
simulations \citep{MC00} nor the DFT-MD simulations \citep{Le00} could
reproduce the results from laser-driven shock measurements. Later
DFT-MD simulations by ~\citet{Bonev2004}, that treated the molecular
phase more accurately, improved the agreement with the gas-gun shock
wave experiments by ~\citet{Nellis1984} below 0.25 Mbar but the
results at higher pressure remained unchanged. The agreement between
the two first-principle simulation methods, PIMC and DFT-MD, is
reasonably good, though not yet perfect. This made it possible to put
together consistent EOS tables for hydrogen~\citep{MC01},
helium~\citep{Mi09}, carbon~\citep{Benedict2014},
nitrogen~\citep{DriverNitrogen2016}, oxygen~\citep{DriverOxygen2015},
water~\citep{DriverMilitzer2012}, neon~\citep{Driver2015}, and most
recently silicon~\citep{MilitzerDriverPRL2015}. In each case, results
from PIMC simulations that are very efficient at high temperature were
combined with DFT-MD results at low temperature.

A major contribution towards resolving the controversy regarding the
deuterium Hugoniot curve came from magnetically driven shock
experiments by~\citet{Kn01}, which favor a maximum shock compression
ratio of $\sim4.3$, broadly consistent with predictions from
first-principles simulations. Later, similar results were reported
from spherically converging shock wave experiments
by~\citet{Belov2002} and by~\citet{Boriskov2003} as well as by planar
shock wave experiments by~\citet{Brygoo2015} that were performed at
the Omega laser facility.  Because of all three new measurements favor
a compression of $\sim4.3$, in keeping with first-principles simulations, one
may regard the controversy around the deuterium Hugoniot curve to be
resolved with satisfactory accuracy. If we adopt this view, however,
then \citet{SC95} EOS would be no longer the best EOS to model giant
planet interiors because it deviates from shock measurements for
$P>0.7\,$Mbar.


\begin{figure}[ht]
  \begin{center}
\noindent\includegraphics[width=16pc]{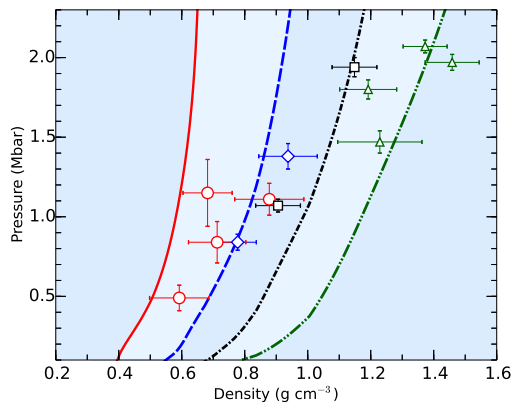}
  \end{center}
  \caption{Comparison of the helium shock Hugoniot curve derived
    DFT-MD simulations (lines, \citet{Mi06,Mi09}) and shock wave
    experiments (symbols, \citet{Brygoo2015}). The curve represent various
    initial densities in the experiment.}
\label{fig:heliumshocks}
\end{figure}

\subsection{Experiments and Simulations of Dense Helium}

Helium has also been studied with first-principles computer
simulations and high-pressure experiments, although to a lesser extent
than hydrogen and deuterium. Early-on diamond anvil cells (DAC) have
been used to explore the solid phases of helium and its melting curve.
\citet{Loubeyre1982} and \citet{Vos1990} measured the melting
temperature of helium 4 from 7.4 to 250~kbar and found it to increase
from 50 to 480~K over this pressure interval. \citet{Vos1990} fit the
melting line with a simple power law,
\begin{equation}
 T_\textrm{melt}= 14.0090\,P^{0.6390},
\end{equation}
with the pressure $P$ in kbar and the melting temperature in Kelvin.
There seems to be some slight deviations from this expression at
higher pressure \citep{Datchi2000} but when extrapolated to the Mbar
pressures, this expression is in good agreement with \textit{ab
  initio} estimates of the melting line
\citep{Lorenzen2009}. \citet{Loubeyre1982} also observed a triple
point at 299~K and 116.5~kbar between the liquid and possibly an fcc
and an hcp solid phase.

The fluid phase of helium has been studied with shock wave
  experiments~\citep{Nellis1984}. In the experiments by
  \citet{Eggert2008}, that reached 2 Mbar, diamond cells were used to
  increase the initial density of the helium samples. Because of the
  complexity and the short time scale of these experiments, the
  particle velocity, $U_p$, was not measured directly. Instead of using
  Eq.~2, 
  the pressures had to be inferred with an impedance matching construction
  \citep{Brygoo2015} that relied on a reference material with known
  properties. For the helium and hydrogen measurements, quartz was
  used as a reference. After the experimental results had been
  published, the quartz shock standard was
  revised~\citep{Knudson2009,Knudson2013} and \citet{Brygoo2015}
  reinterpreted the existing hydrogen and helium measurements. The
  updated helium results are compared to predictions from {\it ab
    initio} simulations in Fig.~\ref{fig:heliumshocks}.

The EOS of the fluid helium has been characterized by analytical free
energy models by \citet{SC95} and slightly improved by more complete
calculations by \citet{Winisdoerffer2005}.  More recently, extensive
first-principles simulations have been performed using PIMC for
temperatures above 10$^5\,$K and DFT-MD for the lower temperatures
\citep{Mi06,Mi09}. The pre-compressed Hugoniot curves predicted by the
\textit{ab initio} calculations are displayed in
Fig.~\ref{fig:heliumshocks}. Overall, there is a good agreement with
experimental results but for small initial densities, the measurements
predict a slightly higher final shock densities than the DFT-MD
simulations.

Unlike hydrogen, helium does not exhibit a molecular phase but one
still expects it to become a metal at high pressure.  This phenomenon
has been investigated with {\it ab initio} simulations, both by
studying the band gap and by computing conductivities and
reflectivities.  The latter quantity can directly measured during
shock experiments. In the solid phase at 0~K, \citet{Khairallah2008}
showed band gap closure occurs at 21.3~g/cm$^3$, or 257 Mbar, using
quantum Monte Carlo calculation that are in very good agreement with
predictions from GW density functional theory (GW-DFT).  Later
calculations with electron-phonon coupling by \citet{Monserat2014}
suggested that the metallization of helium occurs at even higher
pressures. Thus, the metallization pressure of solid helium is at
least one order of magnitude higher than that of hydrogen, which is
expected to metalize at several Mbar.


The temperature effects on the metallization conditions of fluid
helium were studied by \citet{Kowalski2007} with DFT and GW methods.
They showed that the width of the gap depends on the temperature and
estimated a metallization density of 10~g/cm$^3$, which was in
agreement with calculations in the chemical picture
by~\citet{Winisdoerffer2005}.

The first shock wave experiments that measured the reflectivity of
dense helium (~1.5~g/cm$^3$) were reported by \citet{Celliers2010}.
The size of gap was estimated and gap closure was predicted to occur
at a density of 1.9~g/cm$^3$. A re-analysis of the experimental data
by \citet{Soubiran2012}, including the temperature effects on the
helium gap, showed that experimental findings would also be in
agreement with a much higher metallization density of 10~g/cm$^3$ or
above. This implies that unlike hydrogen, pure helium would remain in
an atomic and insulating phase over the entire range of
pressure-temperature conditions in the interior of giant planets.

\section{Hydrogen-Helium Mixtures}
\label{sec:HHe}

\subsection{Experiments at Lower Pressure}

Because of their significance for planetary science, the
hydrogen-helium mixtures have been investigated with various
experimental high-pressure techniques. 
The phase-separation transition in the fluid phase was first studied
with static anvil cell experiments with an optical observation of the
phase transition.  \citet{Streett1973} and \citet{Schouten1985}
reported a partial phase separation into two fluids for pressures up
to 50~kbar over a temperature interval from 26~K to nearly room
temperature.  \citet{Loubeyre1985,Loubeyre1987} used the displacement
of the hydrogen $Q_1$ mode in the Raman spectrum due to the presence
of helium to determine the mixing phase diagram up to 100~kbar and
373~K. A detailed phase diagram based on the available
  experimental data is given in Fig. 9 of \citet{Loubeyre1987}.

\begin{figure}[!hbt]
  \begin{center}
    \noindent\includegraphics[width=19pc]{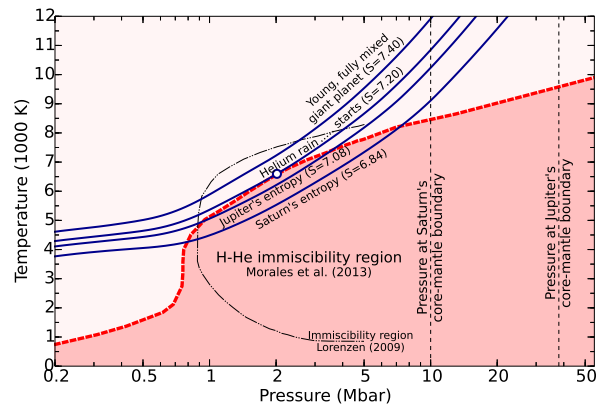}
  \end{center}
  \caption{Hydrogen-helium miscibility diagram. The solid lines show
    DFT-MD adiabats from \citet{MH13} labeled with their entropy in
    units of $k_b$ per electron. The shaded area is the immiscibility
    region calculated by \citet{Morales2013} that we extrapolated
    towards higher pressures. }
  \label{fig:imm}
\end{figure}

The immiscibility of hydrogen and helium depends on temperature,
pressure, and the concentration. For low helium concentration, the
demixing is mostly related to the crystallization of hydrogen. For a
slightly higher helium concentrations, a triple point is observed for
temperatures between 100 and 373 K as the helium concentration is
increased from 0.11 to 0.32. This triple point is an equilibrium
between a hydrogen-rich solid phase and two liquid phases of
intermediate and high helium concentrations.

\citet{Loubeyre1987} also defined a critical point that, for a given
temperature, marks the minimum pressure for any mixtures to show phase
separation. For instance, at 295~K, the critical point is at 51~kbar,
while the triple point is at 62~kbar (see Fig.~9 of \citet{Loubeyre1987}).
While these results represent the best laboratory measurements of
hydrogen-helium mixtures to date, their relevance for the extreme
pressure and temperature conditions in giant planet interiors remains
to be determined.

\subsection{Simulations of Hydrogen-Helium Phase Separation}
\label{sec:phase_sep}

Because of the relevance to giant planet interiors, hydrogen-helium
mixtures have been the subject of various first-principles
studies. Early DFT simulations by~\citet{Klepeis1991} and
~\citet{Pfaffenzeller1995} focused on ground state calculations of solid
hydrogen-helium mixtures because DFT-MD simulation at high temperature could not
yet been performed efficiently. Homogeneous hydrogen-helium mixtures
were also studied with path integral Monte Carlo
simulations~\citep{Mi05} but it has proved challenging to apply this method
below 10$\,$000$\,$K where one expects phase separation in the fluid. 

Once more computer time became available, \citet{Vo07} studied
nonlinear mixing effects in hydrogen-helium mixtures with DFT-MD
simulations. It was demonstrated that, for a given pressure and
temperature, the presence of helium stabilizes the hydrogen molecules
and shifts the molecular-to-metallic transition in hydrogen to higher
pressures. A direct demonstration of the phase separation with
first-principles method was still missing, however. This problem was
elegantly solved with the Gibbs free energy calculations
by~\citep{Morales2009,Morales2013} who determined the full nonideal
entropy of mixing with DFT-MD simulations. The emerging immiscibility
region is shown in Fig.~\ref{fig:imm} and was used in Jupiter interior
model by~\citet{HM16}.

\citet{Lorenzen2009} also performed DFT-MD simulations to derive the
immiscibility region in hydrogen-helium mixtures. While non-ideal
mixing effects of pressure and internal energy were included in the
calculation, only the ideal entropy of mixing was considered. The
resulting immiscibility region, shown in Fig.~\ref{fig:imm}, is
surprisingly close to the results by ~\citet{Morales2013}. Still at 2
Mbar, \citet{Lorenzen2009} overestimate the immiscibility temperature
by 1000 K. Below 3000 K, the \citet{Lorenzen2009} calculations become
invalid because nonideal contributions to the mixing entropy become
very important. Without an explicit calculation, it is difficult to
predict how important nonideal mixing effects are. For instance,
\citet{SoubiranMilitzer2015} found that dense molecular mixtures of
hydrogen and water behaved approximately ideal while
~\citet{WilsonMilitzer2010} demonstrated that one may obtain the wrong
answer in helium rain sequestration calculations unless the nonideal
entropy of mixing is included correctly.

Deep inside the immiscibility region, one can also observe the
phase-separation of the hydrogen and helium directly in the DFT-MD
simulations~\citep{Soubiran2013,Militzer2013}, which provides an
independent confirmation of the results from Gibbs free energy
calculations.

The most common structure assumed for Jupiter's interior is a three layer
model \citep{Guillot2004,SG04,Nettelmann2012,HM16} that consists of a
dense, heavy element rich, central core, along with a metallic inner envelope and a molecular outer
envelope, both comprised predominately of hydrogen and helium. Each layer is assumed to be adiabatic
and of constant composition. The entropy, the helium mass fraction,
$Y$, and mass fraction of heavier elements, $Z$, may differ between
the layers.  Temperature and pressure are typically assumed to be
continuous across layer boundaries.

\section{Current Jupiter Interior Models} 
\label{sec:interior}

One key constraint on Jupiter's atmosphere are the temperature-pressure
measurements by the \textit{Galileo} entry probe shown in
Fig.~\ref{fig:ad}. This can be used as a starting point for the
temperature profile in the outer layer. In addition, the
\textit{Galileo} entry probe provided measurements of $Y$ and $Z$,
which are typically taken to be representative of at
least the outer envelope composition.

Furthermore, models for Jupiter's interior are constrained by the
total mass of the planet, along with the measured gravitational
moments $J_2$ and $J_4$ discussed in section~\ref{sec:space_missions}.
To relate given values of $J_n$ to interior structure, the planet is
assumed to be in hydrostatic equilibrium in the rotating
frame, and a unique barotrope $P=P(\rho)$ is derived from the
equation of state to relating the pressure $P$ and the mass
density $\rho$ for a given entropy and composition.  An acceptable
model must then reproduce the external gravity terms within the uncertainty of
the observations.

\begin{figure}[!hbt]
  \begin{center}
    \noindent\includegraphics[width=16pc]{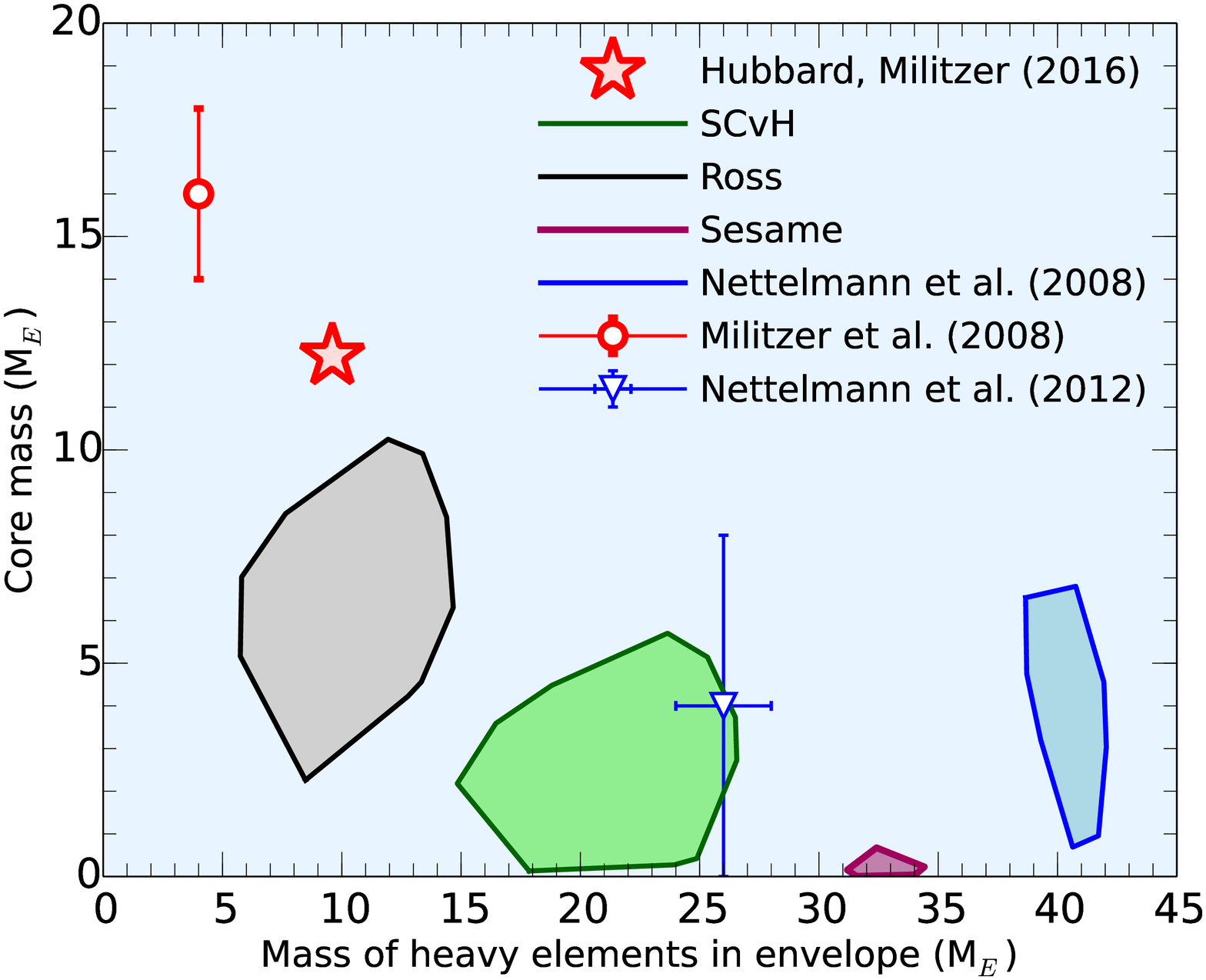}
  \end{center}
  \caption{The mass of the dense central core and amount heavy Z 
    elements in the envelope is compared for different models 
    of Jupiter's interior. Results reported by~\citet{SG04} were 
    complemented with model predictions by~\citet{MHVTB,NHKFRB,Nettelmann2012,HM16}.}
    \label{fig:core}
\end{figure}

\begin{figure}[!hbt]
  \begin{center}
    \noindent\includegraphics[width=17pc]{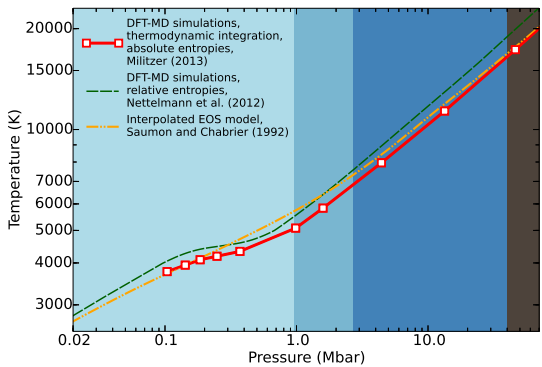}
  \end{center}
  \caption{Different theoretical predictions for Jupiter's interior adiabat. The shaded areas corresponds to the layers
    in figure~\ref{fig:our_Jupiter_model}. }
  \label{fig:jup}
\end{figure}

Determining the $J_n$ of a model requires a self-consistent
calculation of the planet's rotation-induced shape and gravitational
field. While analytic high-order perturbative theories for the zonal
harmonic coefficients have been developed~\citep{ZT1978}, the high
expected precision of the \textit{Juno} model has prompted the
development of more precise non-perturbative, numerical methods for calculating
$J_n$ \citep{Hubbard2013,wisdom2016}. Using this method, \citet{HM16}
suggested that the actual $J_4$ of Jupiter might lie outside of the
error bars reported by \citet{Jacobson2003} in order to be consistent with the
DFT-MD equation of state by~\citet{MH13}.

In Fig.~\ref{fig:core}, we compare the core sizes and amounts of heavy
$Z$ elements in the outer layers that were inferred from model
calculations that assumed various EOSs of hydrogen-helium mixtures.
The large spread of predictions by different equations of state
highlights the importance of an accurate equation of state for the
interpretation of measurements by \textit{Juno} and future missions.
Interior models based on the Sesame \citep{Ke83}, \citet{Ro98}, and
\citet{SC95} EOSs yield core masses of less than 10~$M_\oplus$ and
would thus be inconsistent with the \textit{core accretion} assumption
for the planet's formation \citep{Pollack1996} unless an initial dense
core has been eroded by convection (see
section~\ref{sec:Core_Erosion}). However, these EOSs all yield
substantial core masses between 10 and 25~$M_\oplus$ for Saturn's
interior~\citep{SG04}. This means, if core erosion indeed occurred in
giant planet interiors, it had to be much stronger in Jupiter's
interior than in Saturn. This is not inconceivable since Jupiter is
three time heavier and thus more convective energy would be available to
lift up heavy materials against forces of gravity. 

On the other hand, Jupiter interior models on DFT-MD EOS
by~\citet{MHVTB} and \citet{HM16} predict a larger central core for
Jupiter of 12~$M_\oplus$ or more. The latter model assumed that
helium rain occurred on this planet, which suggests Jupiter's interior
may not be too different from that of Saturn. The DFT-MD based models by~\citet{NHKFRB,Nettelmann2012} predict a
very small central core of 8~$M_\oplus$ or less, which requires further
discussion. 

\begin{figure}[bht]
  \begin{center}
    \noindent\includegraphics[width=16pc]{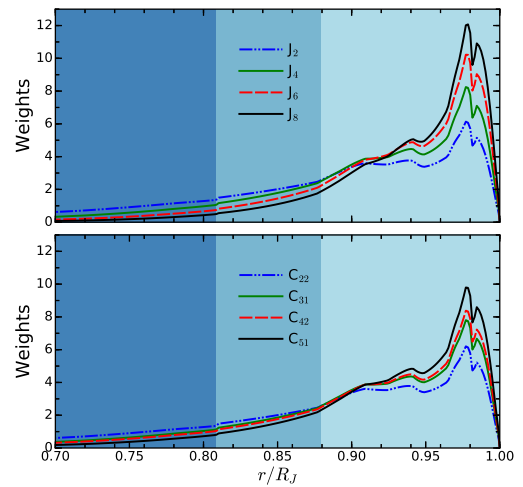}
  \end{center}
  \caption{Contribution of spheroids to Jupiter's external
    gravitational zonal harmonic coefficients (top) and tesseral
    coefficients (bottom) as a function of radius. Weights are normalized such
    that the sum of contributions equals the total number of layers ($N=511$). 
    The shaded areas corresponds to the layers
    in figure~\ref{fig:our_Jupiter_model}. Tesseral moments of the
    same order (i.e. C$_{31}$ and C$_{33}$) have indistinguishable
    radial distributions.}
    \label{fig:weights}
\end{figure}

Figure~\ref{fig:jup} shows that the adiabats are very
different. \citet{MH13} computed absolute entropies. Each point in
Fig.~\ref{fig:jup} was determined with an independent calculation with
the {\it ab initio} thermodynamic integration (TDI) method
\citep{Wijs1998,Alfe2000,Morales2009,WilsonMilitzer2010}. The fact that a smooth curve
emerges demonstrates that the statistical error bars can be controlled
very well with TDI approach~\citep{Militzer2013}. The computed {\it ab
  initio} entropies also agree well with the \citet{SC95} EOS model in
the limit of low and high pressure where one expects this
semi-analytical model to work well since it was designed to
reproduce the EOS of a weakly interacting molecular gas
at low pressures and a two-component plasma in the high-pressure
limit. For intermediate pressures from 0.2 to 20 Mbar,
figure~\ref{fig:jup} shows deviations between the {\it ab initio} TDI
approach~\citep{Militzer2013} and \citet{SC95} EOS model because
little information about the properties of hydrogen near the
molecular-to-metallic transition was available when this model was
constructed. Furthermore, the deviations are seen in a region where
\citet{SC95} interpolated between their descriptions for molecular and
metallic hydrogen. Deviations in this regions are, thus, not unexpected.

It is unusual, however, that the adiabats DFT-MD by
\citet{Nettelmann2012} are significantly higher in temperature than
those by \citet{MH13}.

There are three main reasons for the deviations in the {\it ab initio}
adiabats in Fig.~\ref{fig:core} that need to be considered. (1) Since
~\citet{Nettelmann2012} did not use the TDI technique, absolute
entropies could not be determined. Instead the slope of the adiabat
was inferred from the pressure and internal energies that are
accessible with standard {\it ab initio} simulations. In addition to
the slope, an anchor point at $\sim$0.1 Mbar is needed to start the
computation of the adiabat. Already at this low pressure, the
\citet{Nettelmann2012} abiabat is significantly higher than that
reported by \citet{MH13}. If the anchor point in the
\citet{Nettelmann2012} calculation is chosen differently, the
agreement of the adiabats improves substantially (see Fig. 11 in
\citet{MH13}).

(2) When the slopes of the adiabats are determined from the {\it ab
  initio} pressures and internal energy, $(\partial T/ \partial V)_S=
- T (\partial P / \partial T)_V \, /\, ( \partial E / \partial T )_V
\,$~\citep{Mi09}, one needs $P$ and $E$ on a fine grid of
density-temperature points for interpolation. In practice, one can
only perform a finite number of simulations and the results have
statistical uncertainties.

(3) Finally, ~\citet{Nettelmann2012} computed the EOS for hydrogen and
helium separately and then invoked linear mixing approximation to
characterize the H-He mixture while \citet{MH13} performed fully
interacting simulations on one representative H-He mixture with mixing ratio
($Y$=0.245) and then used the linear mixing approximation only to
perturb around this concentration. However, far outside of the H-He immiscibility
region one expects the linear mixing approximation to be reasonable.

We gathered that the last two points are of lesser importance and
concluded that the main reason for the deviations in
Fig.~\ref{fig:core} was the choice to take an anchor point from the
analytical fluid variational theory \citep{Nettelmann2012}.

The temperature profile is important for models of giant planet
interiors.  For a given pressure, a higher temperature implies a lower
density, which is compensated by a higher fraction of heavy Z elements
when Jupiter interior models are constructed with the goal of matching
the measured values of the planet's gravity field. A higher-than-solar
heavy element fraction would imply that the process, that led to
Jupiter's formation, was more efficient in capturing dust and ice than
gas. Therefore the characterization of H-He adiabats with theoretical
and experimental techniques is important to understand Jupiter's
formation. At the same time, the interaction of heavy Z elements with
dense H-He mixtures needs to be characterized. ~\citet{Soubiran2016}
performed simulations for a variety of heavy elements.  This is the
first study to investigate the properties of multi-component mixtures
of H, He and heavy elements. It shows that the heavy elements slightly
influence the density profile of giant
planets. Fig.~\ref{fig_snapshot} shows one representative snapshot
from a ternary hydrogen-helium-iron mixture.

Giant planet interior models often invoke a different chemical
composition for molecular and metallic layers. The original
justification for introducing this additional degree of freedom was a
first order phase transition between molecular and metallic hydrogen.
This argument is not supported by DFT-MD simulations that show a
smooth transition of properties with increasing
$P$~\citet{Vo07,MHVTB}. More recently it has been proposed that the
separation between layers in Jupiter corresponds to a narrow region of
helium immiscibility \citep{Guillot2004,HM16}. A schematic depiction
of this model is shown in Figure \ref{fig:our_Jupiter_model}. The
precipitation of helium through this layer is expected to lead to an
intrinsic density difference that may inhibit effective convection
between the inner and outer envelope. This may allow Jupiter's deep
interior to be hotter than would be expected for a single layer
convection envelope.  \citet{HM16} identify this region by identifying
the pressures where the present-day adiabat for the outer envelope
intersects H-He immiscibility region from \citet{Morales2013}
(Figure~\ref{fig:ad}). This leads to a prediction of a present-day
helium rain region between $\sim0.81-0.88a$. This model has an
important evolutionary distinction, since the planet's temperature
profile would have initially been above the immiscibility region,
meaning the envelope is initially homogeneous, with the demixing
layer forming and growing as the planet cools.

Figure~\ref{fig:weights} shows the contribution functions for zonal
and tesseral harmonics of degree 2, 4, 6, and 8.  As suggested by
Eq.~\ref{eq:J2ndef}, these functions are progressively more peaked
toward the surface with increasing degree.  There is no significant
direct contribution from the central region where a central core might
exist.  Moreover, the contribution functions overlap substantially,
implying that values of the harmonics for a given density distribution
are strongly correlated.  The mass of a dense central core cannot be
inferred solely from a finite set of harmonic coefficients; instead it
must emerge from a simultaneous fit to all available constraints on
interior structure.  This process is strongly dependent on the
precision of the thermodynamic model used to construct the barotrope.

\section{Discussion of Jupiter Models}
\label{sec:conv}

\subsection{The Adiabatic Assumption}

\begin{figure}[bht]
  \begin{center}
\noindent\includegraphics[width=16pc]{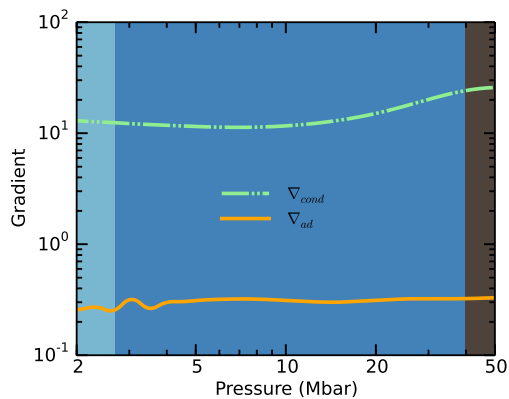}
  \end{center}
\caption{Adiabatic and conductive gradients along a Jupiter-type profile. The model 
and the adiabatic gradient are computed with the SCvH EOS \citep{SC95}. The 
thermal conductivity is estimated with the fully ionized model 
\citep{Potekhin2015}. The shaded areas corresponds to the layers
    in figure~\ref{fig:our_Jupiter_model}.}
\label{fig:gradients}
\end{figure}

The modeling of a giant planet's thermal structure rely on a few simple equations
once spherical symmetry is assumed. The first arises from the conservation of mass,
\begin{equation}\label{eq:masscons}
 \frac{\textrm{d}m}{\textrm{d}r}=4\pi r^2\rho, 
\end{equation}
where $m$ is the mass enclosed in a sphere of radius $r$ and $\rho$
the local density. Second, one assumes hydrostatic equilibrium, which
is a reasonable assumption as no shocks or other dynamical effects of
any significance are present.  This provides a constraint on the
pressure profile:
\begin{equation}\label{eq:hydrostat} 
\frac{\textrm{d}P}{\textrm{d}r}=-\frac{Gm\rho}{r^2},            
\end{equation}
where $P$ is the local pressure and $G$ is the gravitational
constant. In the simplest case, just one additional relationship is
required to determine the interior structure of a planet, which is the
equation of state, $\rho=\rho(P,T)$, of the material in each
layer. This require one to introduce additional assumptions for the
temperature profile unless the pressure is dominated by contributions
from a degenerate electron gas. In general the temperature profile,
$T(r)$, is set by the processes that transfer thermal energy
throughout the planet.  The relationship of the temperature and
pressure profiles can be expressed in the following convenient form,
\begin{equation}
  \label{eq:tempprofile} 
  \frac{\textrm{d}T}{\textrm{d}r}=\frac{T}{P}\frac{\textrm{d}P}{\textrm{d}r}\nabla(T,P) \;,
\end{equation}
with the temperature gradient $\nabla(T,P)\equiv\frac{\textrm{d}\ln
  T}{\textrm{d}\ln P}$, which is set by the energy transfer
mechanisms. In the interiors of planets and stars, there are three
possible mechanisms of energy transfer: radiation, convection, and
conduction. To first approximation, the process that leads to smallest
temperature gradient will be the most efficient one and therefore
dominate the energy transfer in a particular layer.
\begin{equation}\label{eq:gradients}
 \nabla(T,P)=\min(\nabla_\textrm{ad},\nabla_\textrm{rad},\nabla_\textrm{cond}).
\end{equation}


The adiabatic gradient, $\nabla_\textrm{ad}$, corresponds to the
temperature gradient that arises in a convecting material. It is
determined by making the assumption that an advected parcel of fluid
rises fast enough to prevent energy loss through diffusion or
radiation to the surrounding fluid. We also assume that the evolution
is slow enough for the pressure inside and outside of the parcel to
reach an equilibrium. In this case the transformation is quasi-static
and thus isentropic. The resulting adiabatic gradient directly follows
from the equation of state,
\begin{equation}\label{eq:gradad}
 \nabla_\textrm{ad}=\left.\frac{\partial\ln T}{\partial \ln P}\right|_S.
\end{equation}
In the case of a purely conductive layer, the thermal conductivity,
$\lambda$, is the key quantity. From Fourier's law, we know that the
heat flux through the sphere of radius $r$ is given by $F(r)=-\lambda
\frac{\textrm{d}T}{\textrm{d}r}$. The heat flux is related to the
luminosity, $l$, by $F(r)=\frac{l(r)}{4\pi r^2}$. From these relations
and Eq.~\ref{eq:hydrostat}, we find,
\begin{equation}\label{eq:gradcond}
 \nabla_\textrm{cond}=\frac{l P}{4\pi\lambda T G m \rho}.
\end{equation}
In case of a purely radiative layer, under the diffusive
approximation, we find that the radiative gradient is given by
\citep{Rutten2003}:
\begin{equation}\label{eq:gradrad}
 \nabla_\textrm{rad}=\frac{3\kappa l P}{64\pi\sigma T^4 G m },
\end{equation}
with $\kappa$ the opacity of the material and $\sigma$ the Stefan-Boltzmann 
constant.

We can perform an order-of-magnitude calculation to determine which
energy transfer mechanism dominates in different layers of a giant
planet. Radiation is only important in the upper most atmosphere
because the radiative opacity is high in all deeper layers.  For
instance in the metallic region in the deep interiors of giant
planets, free electrons absorb photons very efficiently. In
Fig.~\ref{fig:gradients}, we compare the conductive and adiabatic
gradients for the interior of a giant planet. The adiabatic gradient
was derived from the equation of state by \citet{SC95}. For this
comparison, one can assume that the luminosity is constant throughout
the envelope \citep{Mordasini2012} and use the value of
$8.7\times 10^{-10} L_\odot$.


An approximate value for thermal conductivity can be derived from the
fully ionized model of \citet{Potekhin2015}, which an upper limit for
the conductivity because not all electrons are fully ionized and
scattering processes matter in dense hydrogen-helium
mixtures. Fig. \ref{fig:gradients} underlines that the adiabatic
gradient is nearly two orders of magnitude lower than the conductive
gradient which indicates that the convection is the most efficient
mechanism and that the isentropic approximation is valid as long as
the envelope has a homogeneous composition.

\subsection{Over-turning and Double-Diffusive Convection}

The adiabatic assumption relies on the hypothesis that the envelope is entirely and
efficiently convective. However, as Stevenson \citep{Stevenson1985} pointed out more
than three decades ago, the interior of a giant planet may not be homogeneous in
composition for a number of reasons including late planetesimal accretion, core
erosion and phase separation. In this case, even if the thermal density gradient is
destabilizing, an intrinsic density gradient of composition can become stabilizing
if there is an excess of heavy materials in the warm regions and lighter materials in
the cold regions. This process is termed double-diffusive convection or
semi-convection in the literature. To characterize the behavior of the fluid in the
presence of a gradient of temperature $\nabla_T=\frac{\textrm{d}\ln T}{\textrm{d}\ln
P}$ and a gradient of mean molecular weight (equivalent to a gradient of composition)
$\nabla_\mu=\frac{\textrm{d}\ln \mu}{\textrm{d}\ln P}$, we define a density ratio
\citep{Stern1960,Leconte2012}:
\begin{equation}
 R_\rho=\frac{\alpha_T}{\alpha_\mu}\frac{\nabla_T-\nabla_\textrm{ad}}{\nabla_\mu},
\end{equation}
with $\alpha_T=-\left.\frac{\partial \ln \rho}{\partial \ln T}\right|_{P,\mu}$ and
$\alpha_\mu=\left.\frac{\partial \ln \rho}{\partial \ln \mu}\right|_{P,T}$. In the
case of a destabilizing temperature gradient, $\nabla_T-\nabla_\textrm{ad}>0$, but a
stabilizing compositional gradient, $\nabla_\mu>0$, the convective instability
criterion then becomes $R_\rho^{-1}<1$, called the Ledoux criterion \citep{Ledoux1947}. However, this is
not a simple stability criterion and there are three different observed behaviors (see
Fig. 1 in \citet{Leconte2012} for a schematic diagram).

The stability criterion is given by 
\begin{equation}
 R_\rho^{-1}>\frac{Pr+1}{Pr+\tau},
\end{equation}
with $Pr=\frac{\nu}{\kappa_T}$, the Prandtl number, which is the ratio between the 
kinematic viscosity and the thermal diffusivity, and $\tau=\frac{D}{\kappa_T}$, 
the ratio of the solute particle diffusivity and the thermal diffusivity. If this 
criterion is verified then the layer is stably stratified and no dynamics is 
expected. 

In the case of a marginally stable system,
\begin{equation}
 R_\textrm{min}^{-1}<R_\rho^{-1}\leq\frac{Pr+1}{Pr+\tau},
\end{equation}
a system of oscillatory convection, also called turbulent diffusion, can occur. 

The last case is the layered convection where well-defined layers develop with a 
small-scale convection, when
\begin{equation}
 1<R_\rho^{-1}\leq R_\textrm{min}^{-1}.
\end{equation}
The critical value of $R_\textrm{min}^{-1}$ depends on the properties of the fluid 
under study and its exact value is difficult to estimate 
\citep{Radko2003,Rosenblum2011,Mirouh2012}.

While the oscillatory convection seems irreconcilable with the observed heat flux 
of Jupiter or Saturn \citep{Leconte2013,Nettelmann2015}, layered convection 
could explain the observed properties of the planets. In this layered convection, 
the envelope is divided in successive layers alternating between convective 
sub-cells and diffusive layers of height $h_\textrm{c}$ and $h_\textrm{d}$ 
respectively. In a steady state, the typical timescale for the diffusive and the 
convective layers have to be similar leading to a height ratio:
\begin{equation}\label{eq:heightratio}
 \frac{h_\textrm{d}}{h_\textrm{c}} = Ra_\star^{-1/4},
\end{equation}
where $Ra_\star$ is the modified Rayleigh number defined as
\citep{Leconte2012,Spruit2013}:
\begin{equation}\label{eq:modifRa}
Ra_\star=\frac{\alpha_TgH_P^3}{\kappa_T^2}\alpha^4\left(\nabla_T-\nabla_\textrm{ad}\right),
\end{equation}
with $g$ the local gravitational acceleration, $H_P$ the pressure scale height and
$\alpha=h_\textrm{c}/H_P$. It is interesting to note in Eq. (\ref{eq:heightratio}),
that the heights ratio is proportional to $\kappa_T^{1/2}$. Yet, there is a
significant difference when one uses the fully ionized model \citep{Potekhin2015} or
the \textit{ab initio} calculations \citep{French2012} to estimate the thermal
diffusivity, leading to significant uncertainty in the height ratio. Secondly,
Eq. (\ref{eq:modifRa}) shows that $\alpha$ is a key parameter yet unconstrained
\textit{a priori}. 

\citet{Leconte2012} explored the possible giant planet structures
when layered convection is assumed. If we consider a fixed $\alpha$ throughout the
envelope, it is possible to build models that match all the observable properties of
Jupiter and Saturn, including the heat flux and the gravitational moments. The
permitted values for $\alpha$ for Jupiter range from $3\times 10^{-5}$ to $10^{-2}$
which gives a number of layers ranging between 100 and $3\times10^4$.  For a higher
number of layers, the interior becomes too hot and the average density too low. A
very interesting outcome of the layered convection assumption is that the total
content of heavy element increases from 40~$M_\oplus$ in the adiabatic case to
63~$M_\oplus$ for the most extreme layered convection considered, out of which only
0-0.5~$M_\oplus$ is in a dense central core. Moreover, as the layered convection is
less efficient than large-scale convection in transporting heat, they suggested it as
a possible explanation of the excess luminosity of Saturn \citep{Leconte2013}.

\citet{Leconte2012} did not make any assumption on the origin of the
initial gradient of composition for their layered convection model,
but one possible origin is the phase separation of hydrogen and helium
\citep{Christensen1985}. When considering a layered or
double-diffusive convection, the fluxes of heat and of composition
must be carefully characterized since they affect the evolution of the
planet's interior. To better constrain the evolution of Jupiter in the
case of a H-He phase separation, \citet{Nettelmann2015} used numerical
results of layered and double-diffusive convection simulations with
scaling laws for the heat flux \citep{Mirouh2012,Wood2013}. However,
these scaling laws were computed for the case of miscible fluids and
do not consider the possible influence the phase separation may have
on the fluxes. \citet{Nettelmann2015} based their model on
thermodynamic properties coming from the SCvH EOS \citep{SC95} and on
the phase diagram computations of
\citet{Lorenzen2009,Lorenzen2011}. They observed that to match the
Galileo probe measured helium abundance they needed to modify the
shape of the phase diagram, or the outer layer would become too
depleted in helium. They also found that Jupiter's cooling age was too
long by 1.1~Gyr with He rain and the adiabatic assumption (their
Figure 16).  In the case of a layered convection, they found possible
models that would match the observations and the age of Jupiter with
layers of 100 to 1000~m height.

While important advances have been made using non-adiabatic models, they have also
raised many questions that need to be addressed to understand the formation of
layered convection state. First, the H-He phase diagram must be better
constrained. Likewise, the heat and particle fluxes must be better constrained along
with the influence of the phase separation and corresponding release of latent heat.
Last, the length scale of the convective cells is likely to evolve in time with
merging mechanisms \citep{Spruit2013} that will be competing with the phase
separation. The dynamical effects of these different phenomena could have an
important impact on the predicted evolution of the giant planets.

\subsection{Core Erosion} 
\label{sec:Core_Erosion}

The existence and size of a dense central core of Jupiter is an
outstanding question.  Although a dense central core is a natural
outcome of the preferred \textit{core accretion} hypothesis for
Jupiter's formation \citep{Mizuno1978,Bodenheimer1986,Pollack1996}, it
would not necessarily result in a planet formed by the collapse region
of the disk under self-gravity, e.g. \citep{Boss1997}.  Moreover, it
has been suggested by \citet{Stevenson1982a}, that at high pressures
the stable phases of the high-density materials may become soluble in
liquid hydrogen. As a result, an initial dense core with rocky or icy
composition might erode, with the dense material being redistributed
over a larger region of the planet. This is one mean of forming a
density gradient necessary for a double diffusive region in the
planet's interior \citep{Chabrier2007,Leconte2012,Mirouh2012}. While
many interior models (e.g. \citep{HM16}) require a dense central core
of up to $\sim$20 Earth masses to match the observed gravitational
moments, the model predictions are not sensitive to the radius, or
equivalently density, of this core. Various core compositions are
plausible. It could be a terrestrial iron-rock mixture, a
iron-rock-ice mixture, or be a more diffuse mixture of heavy elements
with liquid hydrogen and helium.

The solubility of various materials have been assessed with DFT-MD
calculations comparing Gibbs free energy of the solution compared to
the separated materials. This is accomplished by performing DFT-MD
simulations with a two-step thermodynamic integration method
\citep{Morales2009,WilsonMilitzer2010}. A number of studies used this
to find the solubility of various analogue planetary materials in
liquid metallic hydrogen. Dissolution was found to be strongly
favorable for both water \citep{Wilson2012a} and iron \citep{Wahl2013}
in the cores of Jupiter and Saturn. Solubility of rocky analogues
$\mathrm{MgO}$ \citep{Wilson2012b} and $\mathrm{SiO}_2$
\citep{Gonzalez2014} are more moderate, but are still predicted to
dissolve in Jupiter's interior. The high-pressure solubility of all
these materials with metallic hydrogen is consistent with their
increasing metallic nature at high pressures. Thus a dense central core of
Jupiter is expected to be presently eroded or eroding. However, the
redistribution of heavy elements by inefficient double-diffusive
convection may be slow compared to evolutionary timescales
\citep{Chabrier2007}, keeping most of the heavy elements confined to a
relatively small region in the planet's deep interior.

\section{Conclusions}

Despite the obvious difficulties to predict how the fields of
planetary science and high-pressure physics will develop in the coming
25 years, a few statements can be made. After {\textit Juno}, we
expect further space missions to visit all four giant planets in our
solar system. Since Uranus and Neptune have not been studied as often
as Jupiter and Saturn, we expect to gather fundamental knowledge about
the history of outer solar system from future ice giant missions. It
would be worthwhile to include entry probes since they provide so much
more detailed information about the atmospheric composition than
remote observations can.

We also expect first-principles computer simulations to become more
accurate. Standard DFT-MD methods may be replaced with quantum Monte
Carlo methods.

As far as high-pressure laboratory experiments are concerned, a direct
measurements of hydrogen-helium phase separation would be of
significant importance for our understanding of gas giant interiors.
We also anticipate that cold, metallic hydrogen will be produced in
static compression experiments at room temperature.



%
%
%
%
%
%
%
\section*{Acknowledgments}

The authors knowledge support from the U.S. National Science
Foundation, from the NASA mission \textit{Juno} and a \textit{Cassini}
data analysis grant. T. Guillot provided valuable comments on
Fig.~\ref{fig:ad}. The data in the figures can be obtained from the
references or by contacting the first author.

\end{article}
%
%
%
%
%
%
%
%


\end{document}